\documentclass{iau}
\usepackage{graphicx}
\usepackage{placeins}

\title[Comparing MOCCA Simulations with Observations using COCOA] 
{MOCCA Code for Star Cluster Simulation: Comparison with Optical Observations using COCOA}

\author[Abbas Askar, Mirek Giersz, Wojciech Pych, Arkadiusz Olech
\& Arkadiusz Hypki]   
{Abbas Askar$^1$, Mirek Giersz$^1$, Wojciech Pych$^1$, Arkadiusz Olech$^1$ \and Arkadiusz Hypki$^1$$^,$$^2$}

\affiliation{$^1$Nicolaus Copernicus Astronomical Centre, Polish Academy of Sciences,\\
ul. Bartycka 18, 00-716 Warsaw, Poland \\ Emails: {\tt askar@camk.edu.pl, mig@camk.edu.pl, pych@camk.edu.pl, olech@camk.edu.pl, ahypki@camk.edu.pl} \\[\affilskip]
$^2$Leiden Observatory, Leiden University, P.O. Box 9513, 2300 RA Leiden, The Netherlands}

\pubyear{2014}
\volume{312}  
\pagerange{}
\jname{Star Clusters and Black Holes in Galaxies Across Cosmic Time}
\editors{A.C. Editor, B.D. Editor \& C.E. Editor, eds.}
\begin{document}

\maketitle

\begin{abstract}
We introduce and present preliminary results from COCOA (Cluster simulatiOn Comparison with ObservAtions) code for a star cluster after 12 Gyrs of evolution simulated using the MOCCA code. The COCOA code is being developed to quickly compare results of numerical simulations of star clusters with observational data. We use COCOA to obtain parameters of the projected cluster model. For comparison, a FITS file of the projected cluster was provided to observers so that they could use their observational methods and techniques to obtain cluster parameters. The results show that the similarity of cluster parameters obtained through numerical simulations and observations depends significantly on the quality of observational data and photometric accuracy.

\end{abstract}

\firstsection 
\section{Introduction}

Due to advancements in computational technology over the past couple of decades, there has been extensive work done in modelling the dynamical evolution of star clusters using numerical simulation codes like direct N-body and Monte Carlo codes. In order to be able to get feedback from observers and vice versa, we need to extend numerical simulations for direct comparisons with observations. For this purpose, we are actively developing the COCOA (Cluster simulatiOn Comparison with ObservAtions) code that can create observational data from the output of numerical simulations of star clusters. COCOA has been developed in Python and it combines various tools to enable quick comparisons between simulation and observational data. As an input, the code uses the extended snapshot from the MOCCA code (Hypki \& Giersz 2013; Giersz et al. 2013). We are working on developing additional features in COCOA. The current version of the code can:

\begin{itemize}
\item Project numerical data from star cluster simulations on to the plane of the sky;
\item Give complete projected snapshot with magnitudes and details of binary systems;
\item Simulate observations of stars in binary systems inside star clusters;
\item Generate FITS file and photometric data from projected snapshot;
\item Compute surface brightness and velocity dispersion profiles (using either projected center of mass velocities or velocities of individual stars in binary systems); 
\item Fit projection to the King model and determine cluster parameters
\end{itemize}

In these proceedings, we present initial results from COCOA for a test star cluster after 12 Gyrs of evolution, simulated using MOCCA. We determined the surface brightness, velocity dispersion profiles and cluster parameters using COCOA. For comparison, a FITS file of the projection was provided to observers so that they could use their observational methods and techniques to obtain cluster photometry.
\section{Comparing MOCCA-COCOA Results \& Observational Data}
The test star cluster evolved using MOCCA for 12 Gyrs was tidally underfilled, had a very high primordial binary fraction (95\%) and initially contained 553k objects. Using the sim2obs tool in COCOA, we generated a FITS image (Figure 1) from the MOCCA snapshot after 12 Gyrs of evolution. The simulated observation of the model cluster was made with a 2.5m optical telescope and the distance to the cluster was set to 5 kpc. The FITS files (2$\times$2 mosaic of the cluster) generated by sim2obs were used by observers to obtain photometric data for 49253 objects extending up to a radius of 465 arcsec. We compared the the surface brightness profile from the photometry done by the observers with the SBP from MOCCA-COCOA results (Figure 1). Photometric data from observers was limited and the observational SBP extends to a smaller radius compared to all the data from MOCCA simulations. For this comparison, we limited the MOCCA-COCOA projected data to $\sim$700 arcsec. After photometric corrections, the SBP from the observational data and projected simulation data match closely. With COCOA we are able to fit the SBP to the King Model. Table 1 shows a comparison of cluster parameters obtained from the SBP and King fit for projected simulation and observation data.
\begin{figure}[!htb]
\begin{center}
 \includegraphics[scale=0.315]{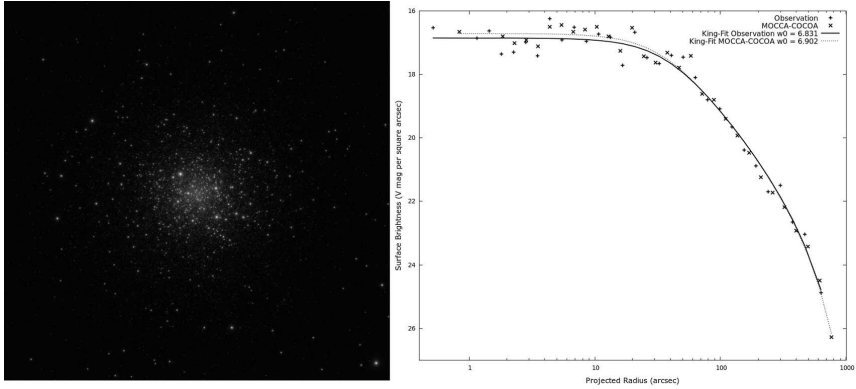} 
 \caption{The left panel shows the grayscale synthetic image of the model cluster created using COCOA. The right panel shows the surface brightness profile and King fit of the cluster using projected MOCCA-COCOA data and observational data.}
   \label{fig1}
\end{center}
\end{figure}
\FloatBarrier
\begin{table}[htbp]
  \begin{center}
  \caption{Comparison of cluster parameters obtained through projected simulation and observational data. These results may vary with different projections of simulation data.}
  \label{tab1}
 {\scriptsize
  \begin{tabular}{|l|c|c|c|}\hline 
{\bf Data} & {\bf King Scale Radius (pc)} & {\bf King Parameter $W_{0}$} & {\bf Half-light Radius (pc)} \\ \hline
MOCCA-COCOA & $0.91$ & $6.90$ & $2.06$\\ \hline
Observation & $0.98$ & $6.83$ & $2.07$\\ \hline
  \end{tabular}
  }
 \end{center}
\end{table}
\section{Acknowledgements}
AA, MG and AH were partially supported by the Polish National Science Centre through the grant DEC-2012/07/B/ST9/04412.

\end{document}